# Characterizing Discourse Group Roles in Inquiry-based University Science Labs

Tong Wan[*]

Department of Physics, University of Central Florida, 4111 Libra Drive, Orlando, FL 32816, USA

Juliette Pimbert

Department of Mechanical and Aerospace Engineering, University of Central Florida, 12760 Pegasus Drive, Orlando, FL 32816, USA

Reshawna L. Chapple

School of Social Work, University of Central Florida, 12805 Pegasus Drive Orlando, FL 32816, USA

Ying Cao

School of Education and Child Development, Drury University, 900 N. Benton Ave. Springfield, MO 65802, USA

Pierre-Philippe A. Ouimet

Department of Physics, University of Regina, 3737 Wascana Parkway, Regina, Saskatchewan, S4S 0A2, Canada

**Abstract**

Group work is commonly adopted in university science laboratories. However, student small-group discourse in university science labs is rarely investigated. We aim to bridge the gap in the literature by characterizing student discourse group roles in inquiry-based science labs. The instructional context for the study was a summer program hosted at a private research university in the eastern United States. The program was designed as a bridge program for matriculating students who were first generation and/or deaf or hard-of-hearing (DHH). Accommodations such as interpreters and technology were provided for DHH students. We analyzed 19 students' discourse moves in five lab activities from the video recordings, resulting in a total of 48 student-lab units. We developed codes to describe student discourse moves: *asking a question, proposing an idea, participating in discussion, chatting off-task*, and *talking with instructor*. Through a cluster analysis using the 48 student-lab units on quantified discourse moves, we identified four discourse styles, *High on-task high social, High on-task low social, Low on-task low social*, and *Low on-task high social*. The results show that individual students tend to demonstrate varying discourse styles in different lab activities; students' discourse styles within the same groups tend to be aligned with their group members. By examining group members' discourse styles in mixed-gender groups, we did not observe a difference in engagement level between female and male students. DHH students in mixed hearing ability groups, however, were observed to have a lower level of engagement compared to their non-DHH group members. We discuss possible factors that may have contributed to the observations for genders and students with different hearing abilities. We also provide suggestions for promoting equitable small-group discourse in university science labs.

[*]Corresponding author: tong.wan@ucf.edu

## I. Introduction
### A. Group work in university science labs

Group work is a typical format in university science laboratories, where students work in small groups to conduct experiments, collect data, and perform analysis. Students collaborate and build on each other's ideas to construct scientific knowledge, develop reasoning and experimentation skills, and practice communicating scientific ideas.

Because of its potential benefits, group work is commonly integrated in many student-centered instructional approaches that are implemented beyond the lab setting. These approaches include but are not limited to guided inquiry, peer instruction [1], and project-based learning [2]. Studies have shown that in a collaborative studio setting, a close attention to group dynamics can help instructors provide intermittent support to get students unstuck and develop ideas with their peers [3–6]. Instructional strategies based on the instructor's attention to students' group interactions include making student thinking visible, noticing and re-voicing student ideas, supporting group interactions to include all students, and facilitating sharing resources [7–9].

Despite the overall improvement in student learning that group work may have contributed to, affective risks [10,11] and inequities [12–16] in group work have been documented in the literature. For example, Cooper et al. [11] found that students' anxiety increases when they don't feel comfortable, or they don't have a positive relationship with their work partners. Quinn et al. [13] reported inequitable gender division of labor in inquiry-based labs. Theobald et al. [14] found that students who perceived that they worked with a dominator had decreased performance. Dasgupta et al. showed that female students' verbal participation increased as the proportion of women in small mixed-gender group increased [16].

### B. Student discourse in collaborative science learning

Although it is now a consensus that understanding group dynamics can increase learning and improve learning experience, categorizing specific aspects of group dynamics is still developing. Prior research on students' group roles using observation data in university science labs [13,15] tends to focus on student non-verbal behaviors (e.g., handling equipment, using computer, writing on paper). Studies that explore student small-group discourse in university science lab settings are scarce.

Analyzing student discourse allows researchers to gain insights into how students interact with their group members and contribute to the group's cognitive [9,17,18] and metacognitive development [19–21]. This is because cognition and metacognition processes are made visible through discourse. Cao and Korestky [9], for example, examined students' discourse on shared resources in studios during interactive virtual labs on thermodynamics. They categorized conditions when sharing resources is likely to happen and found that interaction is necessary, and facilitating conditions include describing a discrepancy; requesting feedback; asking a general question; responding to a question by elaborating, rephrasing, or paraphrasing others' ideas; making ideas public; and providing an alternative idea.

In addition to socially distributed cognitive discourse, researchers are also engaged in examining social metacognitive discourse in small groups. As stated by Chiu [20], social metacognition "distributes metacognitive demands among group members, it increases the visibility of one another's metacognition, improves individual cognition, promotes reciprocal scaffolding, and enhances motivation". Specifically in the field of physics education research, Sayre and Irving [19] identified a new element in the physicist speech genre which they coined as brief, embedded, spontaneous metacognitive talk, or BESM talk. In Sayre and Irving's study, students use BESM talk to communicate their expectations in four ways: understanding, confusion, spotting inconsistencies, and generalized expectations.

We situate our study in the literature of studying students' collaborative learning in small groups and contribute to the body of research by categorizing students' group roles through analyzing students' discourse.

### C. Current study

The overarching goal of this study is to characterize group roles students take on during discourse interaction in inquiry-based university science labs. The instructional context was a summer program for matriculated students, among which a significant fraction of the students was Deaf or Hard-of-Hearing (DHH). Through video coding of coarse grain-sized student *discourse moves* (i.e., in-the-moment verbal actions) and clustering on those discourse moves, we aim to gain insights into how students engage in small-group discourse in labs. Specifically, we investigate the following research questions (RQs):

RQ1: How do the students in a summer program engage in small-group discourse in inquiry-based university science labs?
- a. What are the students' small-group discourse styles in inquiry-based university science labs?
- b. Do the individual students tend to use the same or varied styles in different lab activities?
- c. Do the group members' discourse styles tend to be aligned or diverse?

RQ2: Are students from historically under-represented groups in a summer program engaged in equitable small-group discourse in inquiry-based university science labs?
- a. Are the female students engaged in equitable small-group discourse in inquiry-based university science labs?
- b. Are the DHH students engaged in equitable small-group discourse in inquiry-based university science labs?

## II. Methods
### A. Instructional context
#### 1. The IMPRESS program

The context for the current study was the Integrating Metacognitive Practices and Research to Ensure Student Success (IMPRESS) [22] summer program hosted at a private research university in the eastern United States. It was a 10-day bridge program for matriculating students who were first generation and/or DHH. This program was offered yearly at the university from 2014 to 2017.

The main objectives of the IMPRESS program were to engage students in authentic science practice, to facilitate development of a supportive community, and to help the students reflect on the science and themselves in order to strengthen their learning habits and lead them to a stronger future in STEM fields. The program admitted up to 20 students each year. Students worked in small groups of three or four.

In the current study, we focus our analysis on laboratory experiments conducted in small groups during the time span of IMPRESS. Specifically, we analyzed five lab activities from the year 2015. We chose to focus on the year 2015 for three reasons. First, the video data from the years 2015 and 2016 were best documented and cataloged. Second, the ratio of DHH to non-DHH students in 2015 was almost one to one (9 to 11), which allowed us to make a comparison between the two groups of students. The DHH students in 2016, on the other hand, only made up a small fraction of the total participants (4 out of 19). Third, the communication modes used by DHH students in 2015 included both sign language and speech, but all the DHH students in 2016 used speech. The DHH students in 2015 were thus a better representation for the entire DHH student population.

2. Lab activities

In what follows we describe the lab activities the students participated in as part of the program. We considered an activity a *lab* if it required experimental design or model construction, data collection, and data analysis. If an activity only required an observation of a phenomenon, it was not included in the analysis. Additionally, all the activities included occurred in an indoor lab setting; the outdoor activities were not included.

The overall goal of the IMPRESS lab activities is for the students to gain a better understanding of the greenhouse effect. In other words, to see how increasing levels of $CO_2$ in the Earth's atmosphere "traps" heat. In Lab 1, students were tasked with building a model of the Earth and of Earth's atmosphere. While building this initial model, students were given minimal instructions and allowed to use any of a number of materials provided. When the model had been constructed, the students then took it through "day" and "night" cycles using a heat lamp as a stand in for the sun. During this process they took temperature measurements.

The students were also asked to describe what their expectations were for how the temperature of the model should behave given what they already knew about the greenhouse effect. The results of the temperature measurements were then compared to their expectations.

Following this initial model building, the students were taken through three tailored testing experiments in Labs 2-4 that focus on different aspects of the ground-atmosphere system, namely thermal absorption, the albedo effect, and the greenhouse effect. Students tested the effects of different materials on the temperature of the model. When these activities had been completed, they were asked to build a new model of the ground-atmosphere system that incorporates what they had learned. In Lab 5, students built an apparatus to collect and measure the amount of $CO_2$ released in combustion. A detailed description of the lab activities can be found in Appendix A.

The approximate duration, estimated time allocated for data collection, and the main objectives for each lab are shown in Table 1. When students were not collecting data, they were either constructing their physical model or analyzing/interpreting the data collected. We note that students spent varying amounts of time collecting data (and also varying amounts of time building models and analyzing/interpreting data) in different labs, which could be a contributing factor for the types and/or amount of students' discourse moves.

Table 1. Approximate duration, estimated time allotted for data collection, and main objectives of each lab.

|  | Approximate Duration | Estimated time allocated for data collection | Main objectives |
|---|---|---|---|
| Lab 1 | 90 min | 40 min | To build an initial model for Earth and Earth atmosphere |
| Lab 2 | 120 min | 120 min | To test thermal absorption |
| Lab 3 | 150 min | 120 min | To test the albedo effect |
| Lab 4 | 120 min | 80 min | To test the greenhouse effect and build a new model for Earth and Earth atmosphere |
| Lab 5 | 60 min | Not specified | To build an apparatus to collect and measure the amount of $CO_2$ released in combustion |

**B.     Video data and participants**

In the year 2015, 20 students participated in the program. Among all the participants, 10 were self-identified female and nine were self-identified DHH, who were provided accommodations such as interpreters and computers. Students were divided into five groups, each with four members. The program activities were facilitated by a lead teacher and two teaching assistants (TAs), who were participants from the previous year. The teacher and TAs used questioning techniques to facilitate metacognitive discussion.

Most of the program activities, including the first four labs, took place in the IMPRESS classroom and the last lab was conducted in a chemistry laboratory. Students were self-grouped during the lab activities. They worked in the same groups on labs 1 and 2. At the beginning of Lab 3, students were told to work with someone they had not worked with before. They were again self-grouped. The groups then remained the same for the rest of the labs.

A consent form was given to the students. All the students being recorded consented to participation in the research. All the recorded groups had all the members being part of the research.

For each lab, 3-4 groups were recorded by the cameras attached to the tables with a fixed angle, resulting in 17 recorded group-lab units. Five of the group-lab units were missing key information and therefore were not coded. Two units were missing a significant fraction (about two-fifths) of the lab period; two were missing a group member in the recordings (a member was absent or not captured by the camera); in another unit where no interpreter was present, a DHH student had most of his body blocked in view by a computer, which made the signing not fully visible. We coded the groups only if all the members could be analyzed because we aim to characterize students' discourse group roles. If a member was missing from the analysis, the results would not be comparable to those of groups with all four members included.

The 12 units with complete information were coded. As a result, a total of 19 students (out of the 20 participants in the program) were included in the analysis, with 8 being DHH and 11 being non-DHH, 10 being female and 9 being male.

The group compositions with women and DHH students labeled are shown in Table 2. Some of the DHH students used speech to communicate, and the others used sign language with or without speaking simultaneously. For some groups, an interpreter was present at the table to help interpret when a DHH student signed. Two groups had times when no interpreter was present; author R.L.C., who is fluent in American Sign Language (ASL), transcribed students' sign language based on her interpretation. One of the coders (T.W.) and R.L.C. watched some of the videos together and discussed the transcript to make sure the coder interpreted the transcript accurately.

Table 2. Group compositions with pseudonyms and the labs coded. Female (F) and DHH students with their modes of communication [41] are indicated in the parentheses.

| | Group composition | Lab 1 | Lab 2 | Lab 3 | Lab 4 | Lab 5 |
|---|---|---|---|---|---|---|
| Initial groups | Arya (F)<br>Brittany (F)<br>Daniel<br>Pat (F, DHH, Speaking) | ✓ | | | | |
| | Brock<br>Herb<br>Jakob<br>Justin | ✓ | ✓ | | | |
| | Adam (DHH, Signing and simultaneous speaking)<br>Ashley (F, DHH, Signing)<br>Jessica (F, DHH, Signing and Speaking)<br>Taylor (F, DHH, Signing) | ✓ | | | | |
| New groups | Brittany (F)<br>Jessica (F, DHH, Signing and Speaking)<br>Justin<br>Pat (F, DHH, Speaking) | | | ✓ | ✓ | ✓ |
| | BJ (DHH, Signing and simultaneous speaking)<br>Brett<br>Herb<br>Jakob | | | ✓ | ✓ | |
| | Ashley (F, DHH, Signing)<br>Grace (F)<br>Jill (F, DHH, Speaking)<br>Sara (F) | | | ✓ | ✓ | |
| | Arya (F)<br>Daniel<br>Jack (DHH, Signing)<br>Tasha (F) | | | | | ✓ |

It is worth noting that the mixed groups of gender tend to have the majority of the members being female; 5 out of 6 units had 3 members being female, and the sixth unit had an equal number of male and female. Out of the 6 single-gendered units; 4 were all male and 2 were all female.

The DHH students in the mixed groups of hearing ability, on the other hand, tend to have slightly lower representations; 4 out of 9 units had only one DHH student, and 5 units had an equal number of DHH and non-DHH students. Additionally, 2 units had no DHH students, and 1 unit had 4 DHH students.

### C.     Development of codes for discourse moves in labs

Author T.W. developed the initial definitions of the codes that describe students' discourse moves in science labs. Authors T.W. and Y.C. first independently coded 10-min worth of videos for all four students in a group, documented the instances of discourse moves and took detailed notes. They then compared their coding, resolved inconsistencies, and then refined the code definitions.

The finalized codes for discourse moves are shown in Table 3. The codes for five different types of discourse moves include: *asking a question, proposing an idea, participating in discussion, chatting off-task,* and *talking with instructor*.

Inquiry-based labs often require students to make decisions throughout all phases. When we developed the codes, we intended to capture key discourse moves that help move the group forward towards shared group goals. In the following, we justify our development of codes based on the literature and our understanding of what is important in inquiry-based science labs. Moreover, the codes are grounded in our observations of the video data as they represent the themes emerged from student discourse.

As pointed out by Goos et al. [21], "producing mutually acceptable solution methods and interpretations thus entails reciprocal interaction, which would require students to propose and defend their own ideas, and to ask their peers to clarify and justify any ideas they do not understand." Asking questions, proposing ideas, and participating in discussion are considered key discourse moves that contribute to the decision-making process. By asking a question, a student invites their group members to contribute, which helps make sure the group members come to a consensus. When a student proposes an experiment-related idea, the entire group is given an opportunity to critique and consider alternatives. When a question has been asked or an idea has been proposed, it is expected that the group members respond and engage in meaningful discussion. The code *participating in discussion* includes reciprocal interaction, such as answering questions, providing feedback to other members' ideas, as well as justifying one's own ideas and making observations. We chose to focus on the coarse grain-sized discourse moves as an initial step toward characterizing group roles. We did not further differentiate how a student participates in discussion because it is beyond the scope of the current study.

Chatting off-task can also help group members build rapport with one another in a collaborative learning context. More importantly, prior research has shown several productive functions of off-task talks in relation to collaboration, such as "gaining access to collaboration for self", "recruiting others into the collaboration", and "resisting concentrated authority" [23–26].

Talking with the instructor may be an indicator of a student's willingness to communicate with the instructor. Students could ask the instructor a question or discuss ideas with the instructor. The interaction with an instructor could help the group move forward. It could also demonstrate a student's confidence in their own knowledge and skills.

Table 3. Code definitions and examples for student discourse moves in inquiry-based university labs.

| Code | Definition | Example |
| --- | --- | --- |
| Asking a question | Student askes group member a task-related, non-rhetorical question | *"Now what are your ideas?"* *"How are we going to use this in our model?"* |
| Proposing an idea | Student proposes an idea regarding experimental design/procedures, or data collection or analysis. It may or may not be in the form of a question. Note: when a question is posed, the coder first evaluates whether it concerns an idea about the experimental design or procedure. If it is, the coder codes "proposing an idea"; if not, the coder codes "asking a question" (when it is task-related). | *"Let's use this as a representation of land."* *"We can put holes in that and call it carbon dioxide."* *"Do we want to put the mirror inside the tank?"* |
| Participating in discussion | Student participates in discussion, including answering questions, providing feedback to other's idea, and rhetorical questions | *"Because the matt is too big to fit in and there won't be room for ocean, so-"* *"So that doesn't retain heat but the black does."* |
| Chatting off-task | Student talks with their group on topics not related to the task at hand, such as their personal life, jokes, pronunciation. Can be a statement or a question. | *"That takes some serious skill, drowning in the water."* *"That's why I'm not taking Bio."* |
| Talking with instructor | Student talks with the lead teacher or a TA. It can be a statement or a question. It can either on-task or off-task. | *TA: "Where are you guys so far?"* *Student:" We have the sun, and we are going to put the saran wrap on top."* |

### D. Video coding and inter-rater reliability

Typically, classroom observations of student and instructor behaviors (e.g., COPUS [27] and LOPUS [28]) were coded in 2-min intervals. We decided to reduce the interval duration to 1-min because, through an initial exploration of our video data, we noticed that a discourse move occurred more frequently than every 2 minutes. On the other hand, if the duration was too small, it would be very challenging for the coders to reliably capture all the discourse moves with the given quality of the video recordings. Therefore, we decided to code student discourse moves in 1-min intervals. If a particular type of discourse moves for a student *ever* occurred within a 1-min interval, a result of 1 was documented; otherwise, a 0 was documented. We were able to achieve a reasonable level of inter-rater reliability (IRR) for coding every minute.

The videos were coded by authors T.W. and J.P. in Behavioral Observation Research Interactive Software (BORIS), an event logging software. To investigate IRR, authors T.W. and J.P. independently coded a total of 100-min worth of videos for all group members from two different groups for Lab 1. We then compared the codes and obtained a Cohen's Kappa of 0.71, indicating a substantial agreement (i.e., 0.61-0.80). To explore code-level IRR, we used Gwet's AC1 because Gwet's AC1 is less sensitive to low trait prevalence compared to Cohen's Kappa [29]. In other words, Cohen's Kappa does not accurately reflect the level of agreement when a code has low prevalence, which may be the case for certain codes (e.g., *proposing an idea*). The Gwet's AC1 we obtained for *asking a question, proposing an idea, participating in discussion, chatting off-task,* and *talking with instructor* are 0.89, 0.87, 0.61, 0.72, and 0.87, respectively. The agreement on each code falls in the range of substantial or almost perfect agreement (i.e., 0.81-1.00).

### E. Cluster analysis

To characterize student discourse styles, we conducted a cluster analysis using the five discourse move variables in R. Through clustering, objects that are similar to one another are grouped in the same cluster. The object in our analysis is a student's profile, defined by the prevalences of the five discourse variables in a lab activity. Since we coded 12 group-lab units and each group had four students, the number of student profiles was 48.

To prepare data for a cluster analysis, we first determined the frequency (i.e., number of 1-min intervals) that each variable was coded. To account for different durations for the lab activities, we divided the frequency by the total number of minutes that the activity lasted for.

We used Ward's method, a hierarchical agglomerative clustering technique. This type of clustering technique works in a bottom-up manner. Each student profile was initially a cluster. The profiles that were most similar (i.e., closest in distance in a five-dimensional space determined by the five variables) were combined into a bigger cluster. This was repeated until all the profiles were combined in a single cluster. To determine the optimal number of clusters, we used the "elbow method" [30]. We found the "elbow" at cluster number k = 5 (see Appendix B). The follow-up statistical analysis showed that two of the five clusters were not disguisable from each other. We then combined the two clusters, resulting in a total of four clusters. The four clusters were significantly different from one another in terms of all five variables with large effect. (see Section III).

### F. Statistical analysis

To examine the ways in which the clusters are different, we first used Kruskal-Wallis rank sum test [31]. Effect size was calculated using eta-squared [32] with $0.01 \leq \eta^2 < 0.06$ being small effect, $0.06 \leq \eta^2 < 0.14$ being moderate effect, and $\eta^2 \geq 0.14$ being large effect [33]. We then conducted a post-hoc analysis using Dunn's multiple comparison test [34] with Holm-Bonferroni corrections. All the statistical analyses were conducted in R.

Due to a small sample size, we did not conduct statistical tests to examine the correlation among gender, hearing ability, and discourse moves. Instead, we used a measure, which we named "comparative engagement level" index to compare discourse engagement between gender groups, and between DHH and non-DHH students, respectively, while controlling other variables (See Section III, D and E for detail).

### III. Results

The distribution of fractions of 1-min intervals for all student profiles for each discourse code is shown in Figure 1. The three most prevalent but also widespread codes are *participating in discussion* (M=0.35, SD=0.21), *chatting off-task* (M=0.25, SD=0.19), and *talking with instructor* (M=0.20, SD=0.12). The other two types of discourse move *asking a question* (M=0.10, SD=0.07) and *proposing an idea* (M=0.08, SD=0.09) occurred a lot less frequently.

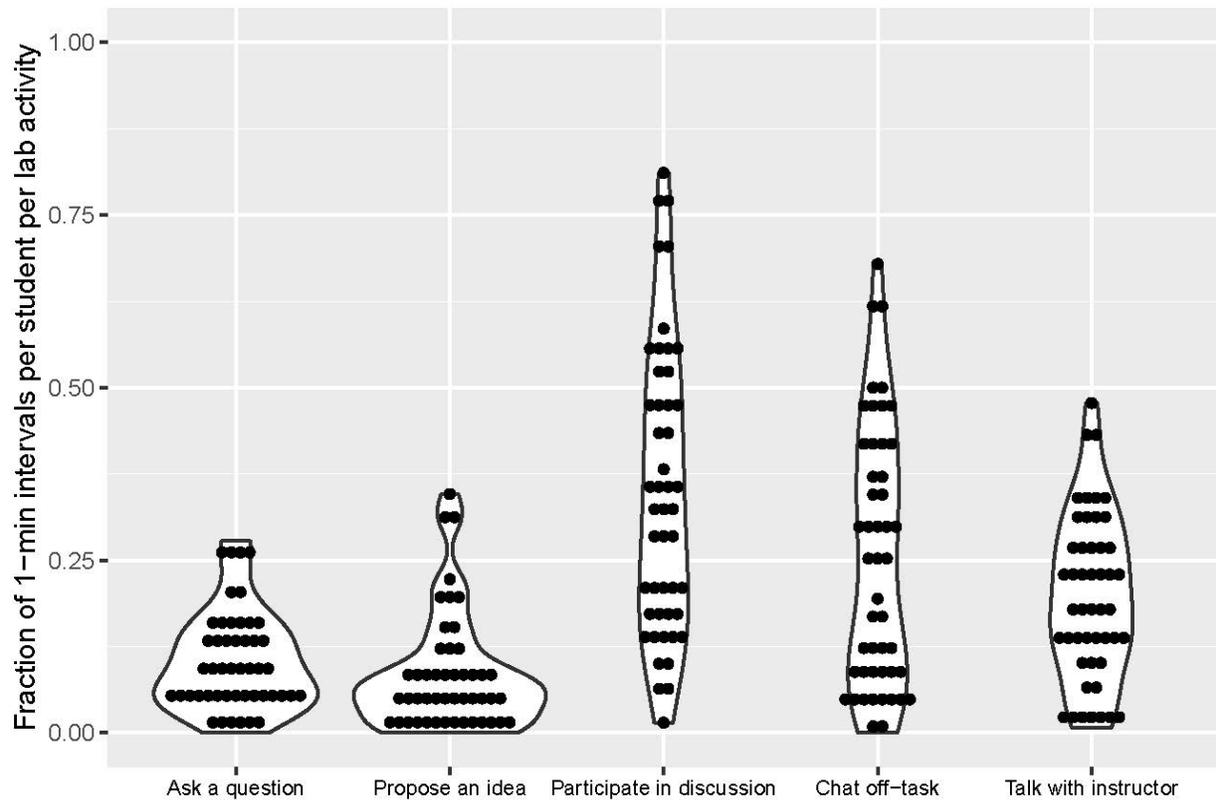

Figure 1. Distribution of fractions of 1-min intervals for all student profiles for each discourse code. Data points represent the student profiles (N = 48) generated from the video coding.

A. RQ1a: characteristics of student discourse styles

The hierarchical structure and cluster memberships of the student discourse profiles are shown in Figure 2. The numbers of profiles in clusters A, B, C, and D are 14, 11, 9, 14, respectively.

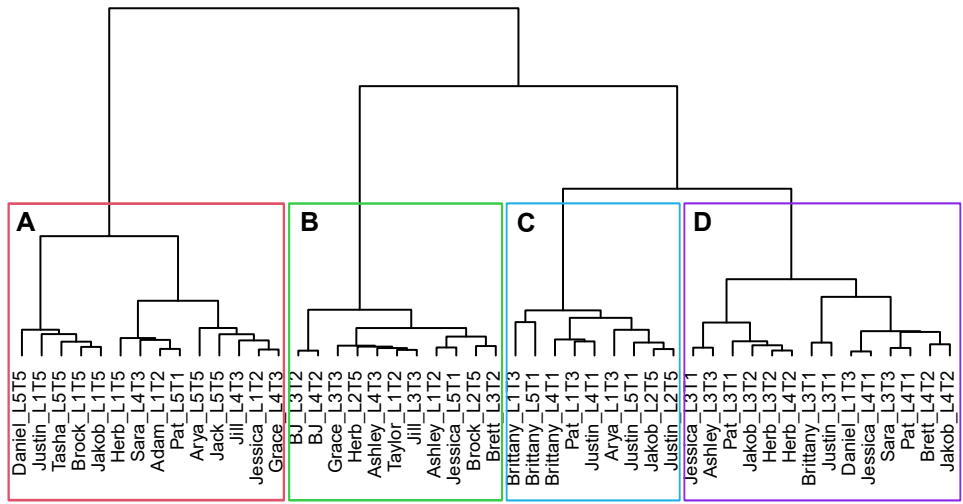

Figure 2. Dendrogram showing the hierarchical structure of the student discourse profiles. Each box represents a cluster of student discourse profiles, each labelled by the student's pseudonym, the lab number, and the table (also the group) number.

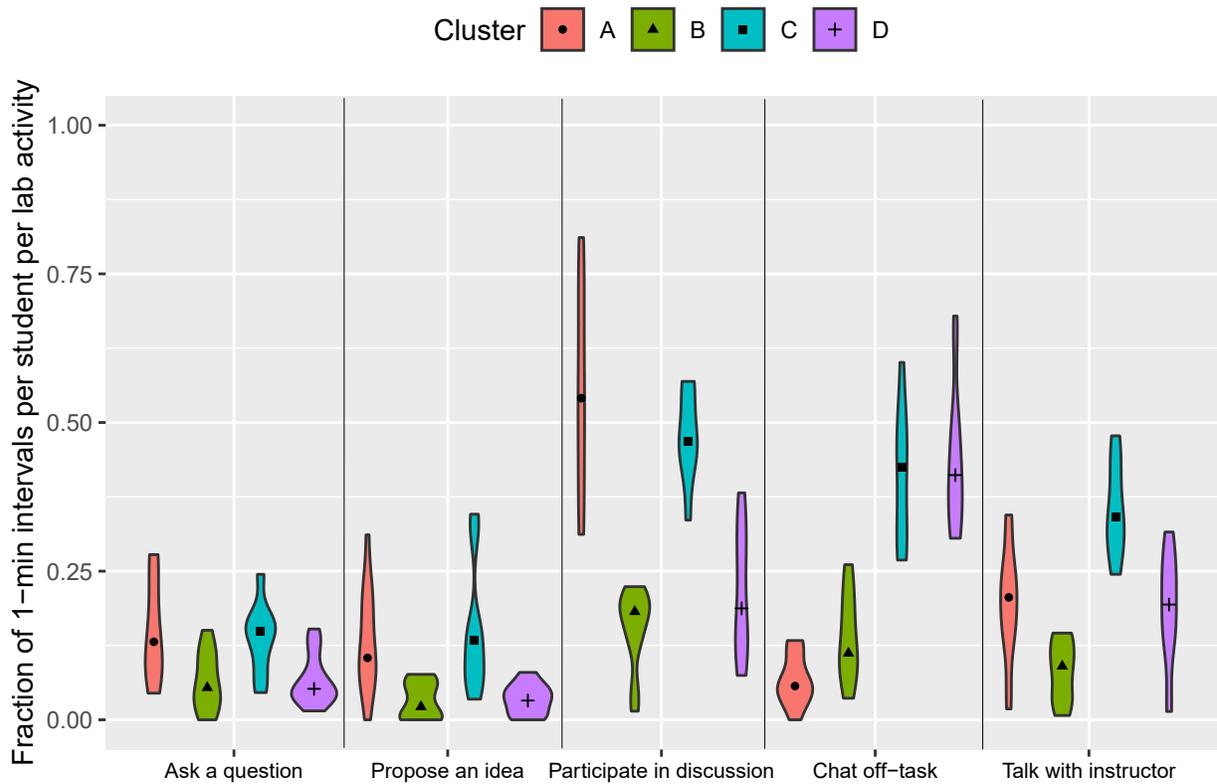

Figure 3. Distribution of fractions of 1-min intervals for each discourse code in each cluster. Points represent median fractions.

The distribution of fractions of 1-min intervals for each discourse code in each cluster is shown in Figure 3. The Kruskal-Wallis tests showed that at least one of the four clusters is distinct from the others in terms of all five discourse codes (p=0.002 for *asking a question*, and p<0.001 for the others) with large effect (see Appendix C, Table 10 for how the clusters are different for each code). Therefore, we use all five discourse codes to describe the characteristics of the four clusters of students' discourse styles. From here on, we refer to the clusters as student discourse styles.

Table 4. Characteristics of student discourse styles in each cluster. Median fractions of 1-min intervals are shown in paratheses.

| Cluster label | Discourse style | Asking a question | Proposing an idea | Participating in discussion | Chatting off-task | Talking with instructor |
|---|---|---|---|---|---|---|
| A | High on-task Low social | High (0.13) | High (0.10) | High (0.54) | Low (0.06) | Medium (0.21) |
| B | Low on-task Low social | Low (0.05) | Low (0.02) | Low (0.18) | Low (0.11) | Low (0.09) |
| C | High on-task High social | High (0.15) | High (0.13) | High (0.47) | High (0.43) | High (0.34) |
| D | Low on-task High social | Low (0.05) | Low (0.03) | Low (0.19) | High (0.41) | Medium (0.19) |

Table 4 shows a summary of the characteristics of student discourse styles. Interestingly, the ways that the four styles are different from one another are identical for the three codes that describe on-task collaboration, *asking a question, proposing an idea,* and *participating in discussion*. That is, styles A and C had higher fractions for these three codes compared to styles B and D. Therefore, we grouped the first three codes into a "on-task" category. Moreover, the code *Chatting off-task* can be used to describe the extent to which students are socializing with their group members. Recall that prior research has suggested that "social" (or off-task talk) has productive functions for collaborative learning.

Since four out of the five codes describe interaction among group members, the four codes (or the two categories, on-task and social) should characterize the main features of the clusters. Therefore, we describe the main characteristics of the four discourse styles in terms of the extents to which students are engaging in on-task collaboration and socializing with their group members. Additionally, we discuss the frequency of talking with instructor for each discourse style.

**Style A: High on-task low social.** Students who used the high on-task low social style asked more questions, proposed more ideas, participated more often in discussions, and talked more frequently with the instructors. They chatted off-task less frequently and had a medium level of interaction with the instructor or TA.

**Style B: Low on-task low social**. Students who used the low on-task low social style not only engaged in on-task collaboration less often, but they also chatted with their group and talked with instructors less often.

**Style C: High on-task high social.** Students who used the high on-task high social style asked more questions, proposed more ideas, and participated more often in discussions. They also chatted off-task more often and talked with instructors most frequently.

**Style D: Low on-task high social**. Students who used the low on-task high social style engaged in on-task collaboration less often, but they chatted more often with their group and had a medium level of interaction with the instructor.

It is worth noting that style C (high-on task high social) is the most desirable among all four styles because it has a high engagement level for all five discourse codes. The second most desirable style is A (high on-task low social), followed by style D (low on-task high social). The least desirable style is B (low on-task low social), which has a low engagement level for all five codes.

### B.     RQ1b: individual students' discourse styles

We found that individual students tend to use a variety of discourse styles in different labs. As shown in Table 5, the number of labs in which an individual student was coded ranged from 1 to 5. For students who were coded in more than one lab, 14 out of 15 students used at least two discourse styles. An individual student could use up to three discourse styles.

Table 5. Distribution of numbers of discourse styles used by individual students.

| Number of labs coded for an individual student | Number of styles observed | Number of students |
|---|---|---|
| 1 | 1 | 4 |
| 2 | 1 | 1 |
| 2 | 2 | 7 |
| 3 | 2 | 1 |
| 4 | 2 | 1 |
| 4 | 3 | 4 |
| 5 | 3 | 1 |

Students' use of a variety of discourse styles appears to be associated with the lab activities they worked on (see Table 6). Two or three groups were coded in all the labs except for Lab 2 where only one group (four male non-DHH students) was coded. Here we discuss the distributions of the discourse styles for labs 1, 3-5. Students' on-task discourse moves started with high prevalences from Lab 1 (50% for *High on-task low social* and 25% for style *High on-task high social* and), and then waned down in Lab 3 (no student used the two styles with high on-task), and finally increased in Lab 5 (63% for style *High on-task low social* and 25% for *High on-task high social*).

Table 6. Number of each discourse style observed in labs 1, 3-5.

|  | Number of students coded | Number of each style observed ||||
|---|---|---|---|---|---|
|  |  | A: High on-task Low social | B: Low on-task Low social | C: High on-task High social | D: Low on-task High social |
| Lab 1 | 12 | 6 (50%) | 2 (17%) | 3 (25%) | 1 (8%) |
| Lab 3 | 12 | 0 | 4 (33%) | 0 | 8 (67%) |
| Lab 4 | 12 | 3 (25%) | 2 (17%) | 2 (17%) | 5 (41%) |
| Lab 5 | 8 | 5 (63%) | 1 (12%) | 2 (25%) | 0 |

Recall that students constructed their initial model of the Earth and Earth's atmosphere in Lab 1, and then they tested the effects of new materials in Labs 2-4. In Lab 4, students also constructed a new model based on the results from all previous testing experiments. During the testing experiments, students repeatedly recorded the temperature for long periods of time (See Table 1). This may be why they used the low on-task styles more prevalently in Lab 3. Additionally, the objective of Lab 5 was to build an apparatus to collect and measure the amount of $CO_2$ from burning material. In this lab, students probably spent more time on designing the experiment than collecting data, which may have contributed to a greater prevalence of the high on-task styles.

Moreover, students regrouped in Lab 3 and then they stayed in the same groups for the rest of the program. Students were asked to work with someone whom they had never worked with before. This may be a reason why the *Low on-task High social* style was much more prevalent than the *Low on-task Low social* style in Lab 3. Students chatted off-task more frequently to socialize with the new groups, which could help them gain access to collaboration.

### C. RQ1c: group members' discourse styles

A student in a group often shared the same discourse style with at least another member. Most (11 out of 12) of the group-lab units coded had only one or two discourse styles observed (see Table 7). There were three instances where all members used the same styles.

Moreover, all four students' discourse styles within a group often shared at least one common feature (i.e., high on-task, low on-task, high social, low social). For example, the group with styles AAAB share low social across all the members, but the members with style A had a higher engagement in on-task discourse. There was only one exception observed: the group with styles ABCC has no feature shared across all four members. However, a common feature was shared with two or three of the members in this group (A and B share low social, and A, C, and C share high on-task).

Table 7. Frequency of number of styles observed within a group.

| Number of styles within a group | Frequency | Observed styles in a group |
| --- | --- | --- |
| 1 | 3 | AAAA, AAAA, DDDD |
| 2 | 8 | AAAB, AABB, BBCC, BBDD, BBDD, BDDD, CCCD, CCDD, |
| 3 | 1 | ABCC |

### D. RQ2a: gender vs discourse

The distribution of fractions of 1-min intervals for each code for different genders is shown in Figure 5. The distributions appear very similar for female and male students.

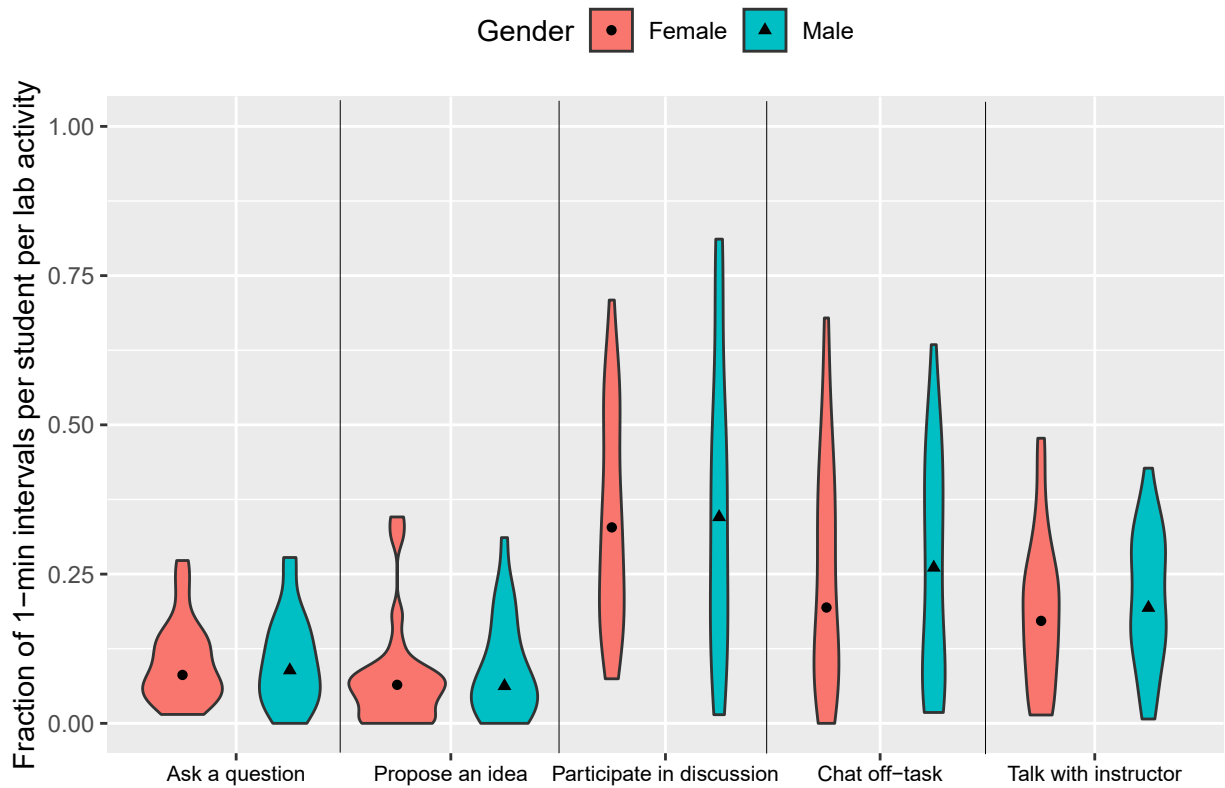

Figure 5. Distribution of fractions of 1-min intervals for each code for different genders. Points represent median fractions.

To explore the correlation between gender and students' use of discourse styles, we examine individual mixed-gender lab-group units and compare the discourse styles of male and female students with the same hearing ability. By examining individual mixed-gender lab-group units we can control the variables of lab activity and group. Moreover, since students worked in groups on the same task, it is more appropriate to compare students' discourse styles within groups. Additionally, we control the variable of hearing ability by comparing male and female students with the same hearing ability.

To quantify the comparison in engagement level between genders, we developed a metric, which we name comparative engagement level (c.e.l.) index. This metric quantifies the frequency that group members in one gender have a higher (or lower) engagement level than group members in another gender have. This frequency is normalized to range from -1 to 1, with a positive value meaning a higher engagement level and a negative value meaning a lower engagement level. If, in all the mixed-gender groups, the engagement level of group members in one gender is higher than that of their group members with another gender, the c.e.l. index would be 1 for the gender with a higher engagement level.

To illustrate how the index was calculated for male students in mixed-gender units, we show the raw and maximum c.e.l. scores for individual units in Table 8. Recall that style C (high-on task high social) has the highest engagement level, followed by style A (high on-task low social), and then style D (low on-task high social), with style B (low on-task low social) having the lowest engagement level (see Sec. III. A. for detail). In unit L1T2 for example, all four members are DHH and we compare the male student's (Adam's) discourse style with the other three female students' styles. Adam's style (high on-task low social) has a higher engagement level than two female (Ashley and Taylor) students' styles (both are low on-task low social) and the same as one female student (Jessica). Adam gets a raw c.e.l. score of +2 because his style has a higher engagement level than two students with a different gender. The maximum raw c.e.l. score Adam could get is 3 because there are three students of a different gender with the same hearing ability. In some other units, there are only 1 or 2 female students with the same hearing ability, then the maximum is 1 or 2, respectively.

We also point out that one could calculate the c.e.l. scores for the female students instead. The total raw c.e.l. for all three female students would have the same absolute value as the male student in the group, but with a different sign. For example, in unit L1T2, the total raw c.e.l. scores for all three females is -2 (Ashely gets -1, Taylor gets -1, and Jessica gets 0). The total maximum raw c.e.l. for female is the same as that for male (the maximum for each female student is 1 and thus the total maximum for three female students is 3).

After we calculate the raw c.e.l. scores for all the male students in all the mixed-gender units, we used the total raw c.e.l. score divided by the total maximum raw c.e.l. score (which can be considered as the normalization factor) to get the c.e.l. index. The calculated c.e.l. index for male students is zero, which shows that no difference in engagement level was observed between male and female students in mixed-gender groups. Moreover, in four of the six units, the male student(s) had a zero raw c.e.l. score. That is, we observed no difference between gender in both individual group level and overall.

Table 8. Calculation of raw c.e.l. for male in mixed-gender units. Each male student in a group was compared with every other female student with the same hearing ability. *Students not included in the comparisons due to different hearing abilities.

| Unit label | Pseudonym | Gender | Hearing ability | Discourse style | Raw c.e.l. score for male | Max raw c.e.l. score for male |
|---|---|---|---|---|---|---|
| L1T2 | Ashley | Female | DHH | B | | |
| | Taylor | Female | DHH | B | | |
| | Jessica | Female | DHH | A | | |
| | Adam | Male | DHH | A | +2 | 3 |
| L1T3 | Arya | Female | non-DHH | C | | |
| | Brittany | Female | non-DHH | C | | |
| | Daniel | Male | non-DHH | D | -2 | 2 |
| | Pat* | Female | DHH | C | | |
| L3T1 | Brittany | Female | non-DHH | D | | |
| | Justin | Male | non-DHH | D | 0 | 1 |
| | Jessica* | Female | DHH | D | | |
| | Pat* | Female | DHH | D | | |
| L4T1 | Brittany | Female | non-DHH | C | | |
| | Justin | Male | non-DHH | C | 0 | 1 |
| | Jessica* | Female | DHH | D | | |
| | Pat* | Female | DHH | D | | |
| L5T1 | Brittany | Female | non-DHH | C | | |
| | Justin | Male | non-DHH | C | 0 | 1 |
| | Jessica | Female | DHH | B | | |
| | Pat | Female | DHH | A | | |
| L5T5 | Arya | Female | non-DHH | A | | |
| | Tasha | Female | non-DHH | A | | |
| | Daniel | Male | non-DHH | A | 0 | 2 |
| | Jack | Male | DHH | A | 0 | 0 |
| | | | | | | |
| | | | | Total | 0 | 10 |
| | | | | Index | 0 | |

### E. RQ2b: hearing ability vs discourse

The distribution of fractions of 1-min intervals for each code for students with different hearing ability is shown in Figure 6. Overall, DHH students tend to have lower fractions across all codes compared to non-DHH students.

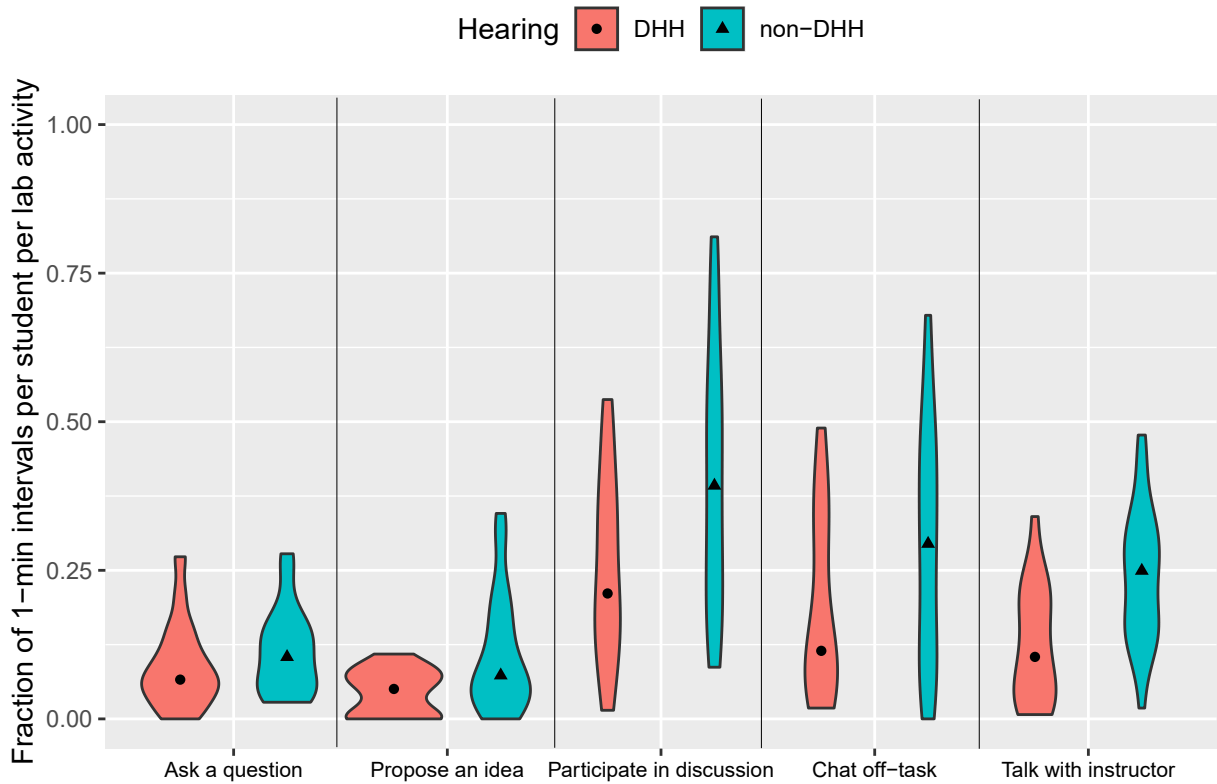

Figure 6. Distribution of fractions of 1-min intervals for each code for students with different hearing abilities. Points represent median fractions.

To control lab activity, group, and gender, we examine individual mixed-hearing-ability lab-group units and compare DHH and non-DHH students with the same gender. Similarly, we calculate the c.e.l. index for DHH students. As shown in Table 9, the total raw c.e.l. score for DHH students is -11, and the maximum is 23. Thus the c.e.l. index can be calculated by dividing the total raw c.e.l. score by the maximum, which gives -0.48. Recall that the c.e.l. index ranges from -1 to 1. The results show that the DHH students overall had a lower level of engagement compared to their non-DHH group members with the same gender.

In 5 (L3T2, L4T2, L4T3, L4T1, L5T1) out of 9 mixed units, DHH students in the groups had lower engagement level compared to their non-DHH group members. In the other 4 units, DHH and non-DHH students had an equal engagement level. By examining the 5 units where a difference between DHH and non-DHH students was observed, we found that DHH students appeared to have a lower level of engagement for on-task collaborations or social interactions compared to non-DHH students within the groups. For example, BJ in unit L4T2 used the low on-task low social style while the other members used the low on-task high social style. Additionally, DHH students also tended to talk with the instructor/TA less often given that the high on-task high social style had the most interactions with instructor/TA and the low on-task low social style had the least interactions.

Table 9. Calculation of raw c.e.l. for DHH students in mixed hearing ability units. DHH student in a group was compared with every other non-DHH student with the same gender. *Students not included in the comparisons due to different genders.

| Unit label | Pseudonym | Gender | Hearing ability | Discourse style | Raw c.e.l. score for DHH | Max raw c.e.l. score for DHH |
|---|---|---|---|---|---|---|
| L1T3 | Arya | Female | non-DHH | C | | |
| | Brittany | Female | non-DHH | C | | |
| | Pat | Female | DHH | C | 0 | 2 |
| | Daniel* | Male | non-DHH | D | | |
| L3T2 | Brett | Male | non-DHH | B | | |
| | Herb | Male | non-DHH | D | | |
| | Jakob | Male | non-DHH | D | | |
| | BJ | Male | DHH | B | -2 | 3 |
| L4T2 | Brett | Male | non-DHH | D | | |
| | Herb | Male | non-DHH | D | | |
| | Jakob | Male | non-DHH | D | | |
| | BJ | Male | DHH | B | -3 | 3 |
| L3T3 | Grace | Female | non-DHH | B | | |
| | Sara | Female | non-DHH | D | | |
| | Ashley | Female | DHH | D | 1 | 2 |
| | Jill | Female | DHH | B | -1 | 2 |
| L4T3 | Grace | Female | non-DHH | A | | |
| | Sara | Female | non-DHH | A | | |
| | Ashley | Female | DHH | B | -2 | 2 |
| | Jill | Female | DHH | A | 0 | 2 |
| L3T1 | Brittany | Female | non-DHH | D | | |
| | Jessica | Female | DHH | D | 0 | 1 |
| | Pat | Female | DHH | D | 0 | 1 |
| | Justin* | Male | non-DHH | D | | |
| L4T1 | Brittany | Female | non-DHH | C | | |
| | Jessica | Female | DHH | D | -1 | 1 |
| | Pat | Female | DHH | D | -1 | 1 |
| | Justin* | Male | non-DHH | C | | |
| L5T1 | Brittany | Female | non-DHH | C | | |
| | Jessica | Female | DHH | B | -1 | 1 |
| | Pat | Female | DHH | A | -1 | 1 |
| | Justin* | Male | non-DHH | C | | |
| L5T5 | Daniel | Male | non-DHH | A | | |
| | Jack | Male | DHH | A | 0 | 1 |
| | Arya * | Female | non-DHH | A | | |
| | Tasha* | Female | non-DHH | A | | |
| | | | | | | |
| | | | | Total | -11 | 23 |
| | | | | Index | -0.48 | |

## IV. Discussion and conclusion

In this study, we characterized students' small-group discourse styles in university science labs. A cluster analysis showed that students demonstrated four different discourse styles, varying significantly from one another in all five discourse variables: *asking a question, proposing an idea, participating in discussion, chatting off-task,* and *talking with instructor*. Characteristics of the discourse styles were then further summarized using the prevalences of engagement in on-task and off-task discourse, namely *High on-task high social, High on-task low social, Low on-task low social,* and *Low on-task high social*.

Students were observed to demonstrate different discourse styles in different lab activities. When the lab provided ample opportunities for decision-making (e.g., a significant portion of time on experimental design was required), more students demonstrated the two high on-task styles (with high prevalences of asking questions, proposing ideas, and participating in discussion). In contrast, when the lab did not require much decision-making (e.g., a significant portion of time on repeated measurements was required), more students demonstrated the two low on-task styles. This was consistent with findings by Quinn et al. [13] who found that students in inquiry-based labs engaged in a variety of activities (e.g., worked on desktop or laptop, handled equipment, and wrote on paper), while students in traditional labs mainly worked on the worksheets. Our findings confirm that the design of lab tasks plays an important role in prompting student engagement.

Moreover, students' use of discourse styles was also influenced by their group members. The results showed that some members in a group, if not all, usually shared the same discourse styles. Additionally, the discourse styles students demonstrated often shared common features with those of their group members. Since students' discourse moves tended to be aligned with their group members, we suggest that interventions for promoting productive group work should focus on the group level rather than the individual level. Students should be given opportunities to reflect and evaluate collectively as a group on how well the group has been functioning. This is in line with the intervention used by Dew et al. [15] who tasked students, as a group, to complete partner agreement and reflection forms in addition to individual reflections.

We did not observe a difference in engagement level in discourse between female and male students in mixed-gender groups. We hypothesize that two potential factors may have contributed to this observation. First, most (five out of six) of the mixed-gender groups had female students in the majority and one group had equal male and female students. Prior research has shown that women's verbal participation in small groups increased as the proportion of women increased [16]. Moreover, for first year students, women in female-majority or female-parity groups were less anxious than those in female-minority groups [16]. Another potential factor is that female and male ratio in the 2015 IMPRESS program was one to one, which was drastically different from many of the university science classrooms (e.g., physics and computer science) where male students usually make up the majority. Studies have showed that the female-to-male ratio in a classroom is positively correlated with female in-class participation [ [35,36]. The findings from the studies mentioned above and the current study together can be interpreted using the "stereotype inoculation model" proposed by Dasgupta [37]. Female peers in class, and in groups particularly, can act like "social vaccines" to "inoculate" against stereotype threat and increase female students' sense of belonging in a group (and class). We suggest that female-majority groups should be prioritized when instructors assign student groups in science classrooms where female students are typically underrepresented.

By examining individual mixed-hearing-ability groups and controlling lab activity and gender, we found that DHH students had lower level of engagement compared to their non-DHH group members. DHH students tended to participate less in on-task conversations and social interactions; they also tended to interact with the instructors less often. The results were consistent with a prior study in high school classrooms that DHH had lower participation than their non-DHH peers despite the availability of technological tools and sign language interpretation [38].

For the DHH students who had a lower level of engagement compared to their non-DHH peers, two students used speech to communicate, one used sign language, and the other two used both sign language and speech. There was no clear evidence that the mode of communication played a role, but it could be due to the small sample size.

One challenge DHH students appeared facing in the lab was that the room could get really loud with all the groups talking simultaneously. Moreover, some group members could have side conversations, which put an additional barrier for DHH students even when an interpreter was present because the simultaneous conversations could make interpretation very challenging. Furthermore, DHH who used sign language to communicate could get tired of watching signing by others. A prior study in the context of undergraduate student research also reported that a DHH was distracted during group meetings when too many conversations were happening simultaneously [39]. One way that could help mitigate this barrier is to raise instructors' and non-DHH students' awareness of DHH students' discourse needs in lab setting.

Another challenge faced by DHH students was confusion masked by the appearance of low motivation, disinterest in the activity or boredom. In our data, there was one group where all four members (Adam, Ashley, Jessica, and Taylor) were DHH, and they used sign language to communicate. Ashley and Taylor (both had low on-task low social style) participated a lot less than Adam and Jessica (both had high on-task low social) did. The low participation of Ashley and Taylor probably resulted from low motivation. During the lab, Ashley expressed that "*I am so done with this*" and "*this is boring.*" Similarly, Taylor wrote in her reflective journal, "*I had fun to play with foil, but this experiment is so boring, I didn't like it or have fun while do it…*" Moreover, Ashley appeared lost in the lab as she wrote in her journal, "*There were really nothing for me to do since I had no clue what the heck, we were doing by creating a model of the atmosphere. I was like what were we supposed to do with this?*" Only about half of the journals scanned were identifiable, and thus we were not able to analyze students' journals to see how prevalent that low motivation was among students with different hearing abilities (or with different genders). Nonetheless, we suggest instructors should increase effort in promoting students' motivation, such as showing support for students and expressing interest in students' ideas and experiences. This would help increase students' sense of belonging, which could in turn help increase student participation.

### V. Limitations and future work

This study has a limitation in its sample size. Five of the 17 group-lab units were excluded from the analysis because they were missing key information (either a significant fraction of the lab period or a group member) in the recordings. Additionally, limited by the recordings, not every student was coded in an equal number of lab activities. Each lab was not coded in equal number of times either. However, our sample had a similar female to male ratio (10 to 9) to the entire program (10 to 10). The DHH to non-DHH was also similar to that in the entire program (9 to 11).

Moreover, this study used coarse grain sized discourse variables to characterize students' discourse styles in university science labs. This allowed us to code the videos relatively efficiently while focusing on the most prominent features in students' discourse styles. However, when coding students' discourse moves, we observed nuanced differences in some of the codes. Analysis addressing those nuances is beyond the scope of this study but can be conducted in future work. For example, the code *participating in discussion* can be broken down into *answering a question, making an observation, providing feedback to others,* and *elaborating on one's idea*. Additionally, a time-sequence log of codes for each group member can be used to examine group dynamics in greater detail. Furthermore, students' discourse moves can be analyzed and characterized at varying levels (type of student interaction, primary intent, and nature of utterance) using the analytical framework proposed by Nennig et al. [34].

The current study investigated students' small-group discourse styles in a science summer program and no gender gap was observed. Future research should investigate discourse styles in different genders in typical university science labs and test the effect of gender ratio in small groups on students' discourse.

Lastly, another future research avenue, in the long term, is to evaluate the effectiveness of group-level interventions on promoting equitable discourse for DHH students. An example of such interventions can include the same elements that were reported in Dew et al. [15], an individual reflection, a partner agreement form, and a partner reflection form. However, these forms need be modified to reflect discourse moves since the original version primarily focuses on what students are handling rather than discourse. Moreover, the challenges faced by DHH students in lab settings should be taken into consideration when developing these interventions.

**Acknowledgement**

We thank the Professional Development for Emerging Education Researchers (PEER) Institute for making the video recordings available for analysis. We also thank Scott Franklin, Eleanor Sayre, and Mary Bridget Kustusch for their support and useful discussions.

**Appendix A: Detailed description of lab activities**

**Lab 1: Initial Model Construction.** On the morning of Day 2, students constructed their initial model of the Earth and Earth's atmosphere using the equipment provided. They measured the temperature of their model during the heating and cooling processes. The students were given minimal guidance and were allowed to proceed as they desired. Following the data collection, students compared the results with their expectations.
**Lab 2: Thermal Absorption.** On the morning of Day 3, students measured the temperature in a model (of the Earth and Earth's atmosphere) with and without a black mat present. The goal was to explore how the heat absorption of a black mat affects the micro-climate of the model.
**Lab 3: Albedo Effect.** On the morning of Day 4, students measured the temperature in a model (of the Earth and Earth's atmosphere) with and without a white foam mat present. The goal was to investigate how the reflection of solar radiation influences the micro-climate of the model.
**Lab 4: Greenhouse Effect – Thermal Absorption of $CO_2$**. On the morning of Day 5, students measured the temperatures of liquids in two flasks, one is pure water and the other contains dissolving tablets that would produce $CO_2$. The goal was to look at how the heating and cooling of the atmosphere differs with and without $CO_2$ in the air. Students were also tasked to revise their model of Earth and Earth's atmosphere incorporating what they had learned from the first four labs.

**Lab 5: Burning Model.** On the morning of Day 7, students carried out a chemistry experiment. The purpose was to build an apparatus that would collect and measure the amount of $CO_2$ released from burning material.

**Appendix B: Elbow plot**

The elbow plot shows the "elbow" at cluster number k = 5 (See Figure 9).

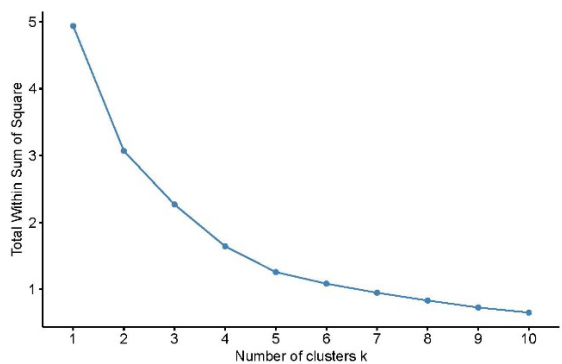

Figure 9. Elbow plot used to determine the optimal number of clusters. The total within sum of square is plotted as a function of the number of clusters.

**Appendix C: Detailed results of Kruskal-Wallis and Dunn's tests**

Table 10. Kruskal-Wallis and Dunn's test results for each discourse code. Results of eta-squared for effect size are also shown.

| Code | $\chi^2(3)$ | $p_{k-w}$ | $\eta^2$ | Ranking by Dunn's test |
|---|---|---|---|---|
| Asking a question | 15.306 | 0.002** | 0.280† | A = C > B = D |
| Proposing an idea | 23.45 | <0.001*** | 0.465† | A = C > B = D |
| Participating in discussion | 33.262 | <0.001*** | 0.688† | A = C > B = D |
| Chatting off-task | 36.982 | <0.001*** | 0.772† | A = B < C = D |
| Talking with instructor | 25.274 | <0.001*** | 0.506† | C > A = D > B |

**p<0.01. ***p<0.001. †Large effect.

**References**

[1] C. H. Crouch and E. Mazur, Peer Instruction: Ten years of experience and results, Am J Phys **69**, 970 (2001).

[2] D. Kokotsaki, V. Menzies, and A. Wiggins, Project-based learning: a review of the literature, Improving Schools **19**, 267 (2016).

[3] M. Koretsky, J. Keeler, J. Ivanovitch, and Y. Cao, The role of pedagogical tools in active learning: a case for sense-making, Int J STEM Educ **5**, (2018).

[4] M. D. Koretsky, Cognitive and social aspects of engagement in active learning, Chem Eng Educ **51**, 198 (2017).

[5] M. D. Koretsky, Towards a stronger covalent bond: Pedagogical change for inclusivity and equity, Chem Eng Educ **52**, 117 (2018).

[6] M. D. Koretsky, Program level curriculum reform at scale: Using studios to flip the classroom, Chem Eng Educ **49**, 47 (2015).

[7] I. S. Horn, *Strength in Numbers: Collaborative Learning in Secondary Mathematics* (Reston, VA: National Council of Teachers of Mathematics, 2012).

[8] M. Windschitl and A. Calabrese Barton, *Rigor and Equity by Design: Locating a Set of Core Teaching Practices for the Science Education Community*, in *Handbook of Research on Teaching*, edited by D. H. Gitomer and C. A. Bell, 5th ed. (AERA Press, Washington, D.C., 2016), pp. 1099–1158.

[9] Y. Cao and M. D. Koretsky, Shared resources: Engineering students' emerging group understanding of thermodynamic work, Journal of Engineering Education **107**, 656 (2018).

[10] L. D. Conlin and R. E. Scherr, Making space to sensemake: epistemic distancing in small group physics discussions, Cogn Instr **36**, 396 (2018).

[11] K. M. Cooper, V. R. Downing, and S. E. Brownell, The influence of active learning practices on student anxiety in large-enrollment college science classrooms, Int J STEM Educ **5**, (2018).

[12] D. Doucette and C. Singh, Making Lab Group Work Equitable and Inclusive, J Coll Sci Teach **52**, 3 (2023).

[13] K. N. Quinn, M. M. Kelley, K. L. Mcgill, E. M. Smith, Z. Whipps, and N. G. Holmes, Group roles in unstructured labs show inequitable gender divide, Phys Rev Phys Educ Res **16**, (2020).

[14] E. J. Theobald, S. L. Eddy, D. Z. Grunspan, B. L. Wiggins, and A. J. Crowe, Student perception of group dynamics predicts individual performance: Comfort and equity matter, PLoS One **12**, (2017).

[15] M. Dew, E. Hunt, V. Perera, J. Perry, G. Ponti, and A. Loveridge, Group dynamics in inquiry-based labs: Gender inequities and the efficacy of partner agreements, Phys Rev Phys Educ Res **20**, (2024).

[16] N. Dasgupta, M. M. M. Scircle, and M. Hunsinger, Female peers in small work groups enhance women's motivation, verbal participation, and career aspirations in engineering, Proc Natl Acad Sci U S A **112**, 4988 (2015).

[17] J. Osborne, Arguing to Learn in Science: The Role of Collaborative, Critical Discourse, Science (1979) **328**, 463 (2010).

[18] S. Kaartinen and K. Kumpulainen, Collaborative inquiry and the construction of explanations in the learning of science, Learn Instr **12**, 189 (2002).

[19] E. C. Sayre and P. W. Irving, Brief, embedded, spontaneous metacognitive talk indicates thinking like a physicist, Physical Review Special Topics - Physics Education Research **11**, (2015).

[20] M. M. Chiu, Social metacognition in groups: Benefits, difficulties, learning, and teaching Computer-assisted Collaborative Learning-Student Learning from Online Discussions View project Discourse Analytics View project, Journal of Education Research **3**, 1 (2009).

[21] M. Goos, P. Galbraith, and P. Renshaw, Socially mediated metacognition: creating collaborative zones of proximal development in small group problem solving, Educational Studies in Mathematics **49**, 193 (2002).

[22] S. V Franklin, E. Hane, M. B. Kustusch, C. Ptak, and E. C. Sayre, Improving Retention Through Metacognition: A Program for Deaf/Hard-of-Hearing and First-Generation STEM College Students, J Coll Sci Teach **48**, 21 (2018).

[23] J. M. Langer-Osuna, E. C. Gargroetzi, R. Chavez, and J. Munson, *Rethinking Loafers: Understanding the Productive Functions of Off-Task Talk During Collaborative Mathematics Problem-Solving*, in *13th International Conference of the Learning Sciences (ICLS)* (2018), pp. 745–751.

[24] C. A. Hass, F. Genz, M. B. Kustusch, P.-P. A. Ouimet, K. Pomian, E. C. Sayre, and J. P. Zwolak, *Studying Community Development: A Network Analytical Approach*, in *2018 Physics Education Research Conference Proceedings* (2018).

[25] P. Hutchison, *Equity and Off-Task Discussion in a Collaborative Small Group*, in *Physics Education Research Conference Proceedings* (American Association of Physics Teachers, 2022), pp. 243–248.

[26] J. M. Langer-Osuna, Productive disruptions: Rethinking the role of off-task interactions in collaborative mathematics learning, Educ Sci (Basel) **8**, (2018).

[27] M. K. Smith, F. H. M. Jones, S. L. Gilbert, and C. E. Wieman, The classroom observation protocol for undergraduate stem (COPUS): A new instrument to characterize university STEM classroom practices, CBE Life Sci Educ **12**, 618 (2013).

[28] J. B. Velasco, A. Knedeisen, D. Xue, T. L. Vickrey, M. Abebe, and M. Stains, Characterizing instructional practices in the laboratory: The laboratory observation protocol for undergraduate STEM, J Chem Educ **93**, 1191 (2016).

[29] N. Wongpakaran, T. Wongpakaran, D. Wedding, and K. L. Gwet, A comparison of Cohen's Kappa and Gwet's AC1 when calculating inter-rater reliability coefficients: a study conducted with personality disorder samples, Medical Research Methodology **13**, (2013).

[30] M. A. Syakur, B. K. Khotimah, E. M. S. Rochman, and B. D. Satoto, *Integration K-Means Clustering Method and Elbow Method for Identification of the Best Customer Profile Cluster*, in *IOP Conference Series: Materials Science and Engineering*, Vol. 336 (Institute of Physics Publishing, 2018).

[31] W. H. Kruskal and W. A. Wallis, Use of Ranks in One-Criterion Variance Analysis, J Am Stat Assoc **47**, 583 (1952).

[32] M. Tomczak and E. Tomczak, The Need to Report Effect Size Estimates Revisited. An Overview of Some Recommended Measures of Effect Size, 2014.

[33] J. M. Maher, J. C. Markey, and D. Ebert-May, The other half of the story: Effect size analysis in quantitative research, CBE Life Sci Educ **12**, 345 (2013).

[34] O. J. Dunn, Multiple Comparisons Using Rank Sums, Technometrics **6**, 241 (1964).

[35] K. Canada and R. Pringle, The Role of Gender in College Classroom Interactions: A Social Context Approach, 1995.

[36] E. G. Bailey, R. F. Greenall, D. M. Baek, C. Morris, N. Nelson, T. M. Quirante, N. S. Rice, S. Rose, and K. R. Williams, Female in-class participation and performance increase with more female peers and/or a female instructor in life sciences courses, CBE Life Sci Educ **19**, 1 (2020).

[37] N. Dasgupta, Ingroup experts and peers as social vaccines who inoculate the self-concept: The stereotype inoculation model, Psychol Inq **22**, 231 (2011).

[38] M. Salehomoum, Inclusion of Signing Deaf or Hard-of-Hearing Students: Factors That Facilitate Versus Challenge Access and Participation, Perspect ASHA Spec Interest Groups **5**, 971 (2020).

[39] A. U. Gehret, J. W. Trussell, and L. V Michel, Approaching Undergraduate Research with Students who are Deaf and Hard-of-Hearing, Journal of Science Education for Students with Disabilities **20**, 20 (2017).

[40] H. T. Nennig, N. E. States, M. T. Montgomery, S. G. Spurgeon, and R. S. Cole, Student interaction discourse moves: characterizing and visualizing student discourse patterns, Disciplinary and Interdisciplinary Science Education Research **5**, (2023).

[41] The DHH students in the study communicated using various communication modalities: speech, American Sign Language (ASL), Signing Exact English (S.E.E.), the combination of spoken-English and sign language (Simultaneous communication or Sim-Com) or spoken- English. DHH student's language use is directly tied to their Deaf Cultural Identity. For example, students who consider themselves culturally Deaf through family or attendance at deafness-orientated schools will be more drawn to using ASL without speech. Students with non-DHH family members and friends may be drawn to using S.E.E signs, Sim-Com, or spoken-English. We do not include the specific sign languages students used in the table because we did not code students' signing directly; instead, we coded the interpretations by the interpreters.